\documentclass[eqsecnum,aps,showpacs,amsmath,twocolumn,superscriptaddress]{revtex4}
\usepackage{epsfig}
\usepackage{color}
\usepackage{graphicx}
\usepackage{amssymb}
\usepackage{algorithm}
\usepackage{algorithmic}
\usepackage{bm}
\usepackage{subfigure}
\usepackage{url}

\begin{document}
\title{Comparison of various methods to extract ringdown frequency from gravitational wave data}

\author{Hiroyuki Nakano}\email{hinakano@law.ryukoku.ac.jp}
\affiliation{Faculty of Law, Ryukoku University, Kyoto 612-8577, Japan}
\author{Tatsuya Narikawa}\email{narikawa@tap.scphys.kyoto-u.ac.jp}
\affiliation{Institute for Cosmic Ray Research, The University of Tokyo,
Chiba 277-8582, Japan}
\affiliation{Department of Physics, Kyoto University, Kyoto 606-8502, Japan}
\author{Ken-ichi Oohara}\email{oohara@astro.sc.niigata-u.ac.jp}
\affiliation{Graduate School of Science and Technology, 
Niigata University, Niigata 950-2181, Japan}
\author{Kazuki Sakai}\email{k-sakai@nagaoka-ct.ac.jp}
\affiliation{Department of Electronic Control Engineering,
National Institute of Technology, Nagaoka College, Niigata 940-8532, Japan}
\author{Hisa-aki Shinkai}\email{hisaaki.shinkai@oit.ac.jp}
\affiliation{Faculty of Information Science \& Technology, 
Osaka Institute of Technology,
Kitayama, Hirakata City, Osaka 573-0196, Japan}
\author{Hirotaka Takahashi}\email{hirotaka@kjs.nagaokaut.ac.jp}
\affiliation{Department of Information and Management Systems Engineering,
Nagaoka University of Technology, Niigata 940-2188, Japan}
\affiliation{Earthquake Research Institute, The University of Tokyo, 
Tokyo 113-0032, Japan}
\author{Takahiro Tanaka}\email{t.tanaka@tap.scphys.kyoto-u.ac.jp}
\affiliation{Department of Physics, Kyoto University, Kyoto 606-8502, Japan}
\affiliation{Center for Gravitational Physics, Yukawa Institute for Theoretical Physics, Kyoto University, Kyoto 606-8502, Japan}
\author{Nami Uchikata}\email{uchikata@astro.sc.niigata-u.ac.jp}
\affiliation{Institute for Cosmic Ray Research, The University of Tokyo,
Chiba 277-8582, Japan}
\affiliation{Graduate School of Science and Technology, 
Niigata University, Niigata 950-2181, Japan}
\author{Shun Yamamoto}
\affiliation{Faculty of Information Science \& Technology, 
Osaka Institute of Technology,
Kitayama, Hirakata City, Osaka 573-0196, Japan}
\author{Takahiro S. Yamamoto}\email{yamamoto@tap.scphys.kyoto-u.ac.jp}
\affiliation{Department of Physics, Kyoto University, Kyoto 606-8502, Japan}

\date{\today}

\begin{abstract}
The ringdown part of gravitational waves in the final stage of merger of compact objects
tells us the nature of strong gravity and hence can be used for testing theories of gravity.
The ringdown waveform, however, fades out in a very short time with a few cycles, 
and hence it is challenging to extract the ringdown frequency and its damping time scale. 
We here propose to build up a suite of mock data of gravitational waves to compare the 
performance of various approaches developed to detect the dominant quasi-normal mode from an excited black hole after merger. 
In this paper we present our initial results of comparisons of the following five methods; 
(1) plain matched filtering with ringdown part (MF-R) method, 
(2) matched filtering with both merger and ringdown parts (MF-MR) method,
(3) Hilbert-Huang transformation (HHT) method, 
(4) autoregressive modeling (AR) method, and 
(5) neural network (NN) method. 
After comparing the performances of these methods, we discuss our future projects. 

\end{abstract}

\maketitle

\section{Introduction}

Both in the year 2016 and 2017, physics community was really got excited by the reports of direct detections of gravitational waves (GWs) by LIGO/Virgo collaborations\cite{GW150914,GW151226,GW170104,GW170608,GW170814,GW170817}. 
The direct detections definitely prove the correctness of the fundamental idea of general relativity (GR), together with that of the direction of efforts of theoretical and experimental researches of gravity. 

LIGO/Virgo collaboration so far performed their observations twice [Observing runs 1 (O1), Sep. 12, 2015 -- Jan. 19, 2016 (48.6 days); O2, Nov. 30, 2016 -- Aug. 25, 2017 (118 days)], and officially
reported~\cite{1811.12907,1811.12940}
that they detected eleven events;  
ten binary black hole (BH) events 
and one event of the merger of binary neutron stars (GW170817~\cite{GW170817}). 
Both types of sources gave us certain advances not only to physics, but also to astrophysics.  

In late 2019, another ground-based GW detector, KAGRA, will join the network of GW observation~\cite{KAGRA_NatureAstro,KAGRA_phase1,LVK_scenario}. 
This will make the source localization more precise, and we also expect to detect the polarization of GWs for each event.    
By accumulating observations, we will be able to investigate new aspects of physics and astronomy, 
such as the distribution of binary parameters, 
formation history of binaries, 
equation of state of the nuclear matter, 
and cosmological parameters. 

Among such possibilities, our interest lies in testing various gravity theories.  
GR has passed all the tests in the past century, and nobody doubts gravity is basically described by GR. 
However, almost of the tests so far have been performed in the weak gravity regime (tests around the Earth, in the Solar System, or using binary pulsars)~\cite{Will_gravitytest}, 
and we still require tests in the strong gravity regime, which is relevant to describe, say, BH mergers. 
Observations of GWs from binary BHs will enable us to test the gravity theories in this extreme regime. 

The previous detections of BH mergers have shown that the inspiral phase (pre-merger phase) is well described by post-Newtonian approximation. 
But it is not entirely clear whether or not the ringdown phase (post-merger phase), which is expected to be well described by BH perturbation theory, was detected in the GW data.
This is because identifying ringdown modes of a BH from noisy data is a quite challenging task for data analysis. 
Ringdown modes decay quite rapidly for a typical BH described by GR. 
For example, for a typical BH formed after merger with the total mass $M=60M_\odot$ and the angular momentum (normalized Kerr parameter) $\chi=0.75$, 
the dominant ringdown mode ($\ell=m=2$) has the characteristic frequency $f_{\rm R}=300$\,Hz 
and the damping time $\tau=3.7$\,msec, which indicates that the amplitude is reduced to 
about 40\% after one-cycle of oscillation. 

One way to give a clear evidence of detection of the ringdown mode would be just to improve the 
detector sensitivity. However, it will also give a similar impact if we can improve the significance 
of detection by implementing an optimized data analysis method. 
There have been already several technical proposals of method to identify the ringdown mode (see, e.g., Refs.~\cite{BertiCardosoCardosoCavaglia2007,BCGS2007,DamourNagar2014,LondonShoemakerHearly2014} or reviews, e.g., Refs.~\cite{Bertietal2009,CardosoGualtieri2016}), but 
we think that fair comparison of the performance of different methods has not been presented yet. 
To find out the optimal method, we organize mock-data tests.  
The idea is to extract the information of the ringdown part [its frequency $f_{\rm R}$ and damping time $\tau$ 
(imaginary part of frequency $f_{\rm I}$ or quality factor $Q$)] 
independently of the other parts of the waveform.  
In order not to allow to identify the properties of mergers from the inspiral waveform using relations valid in GR, 
we prepare a set of blind data each of which has randomly chosen $f_{\rm R}$ and $f_{\rm I}$ different from the GR predicted value (see Sec.~\ref{section2.1}).

In general, the ringdown part includes not only the dominant mode but also subdominant modes.
The analysis of various modes (BH spectroscopy) is also important for test of gravity theories
(e.g., Ref.~\cite{Bertietal2018}).
But, the amplitudes of these subdominant modes are small compared with the one of the dominant
mode~\cite{LondonShoemakerHearly2014}.
Thus, in this work, we focus on the case existing only single mode.

We present our comparisons of the following five methods; 
(1) matched filtering with ringdown part (MF-R) method, 
(2) matched filtering with both merger and ringdown parts (MF-MR) method,
(3) Hilbert-Huang transformation (HHT) method, 
(4) autoregressive modeling (AR) method, and 
(5) neural network (NN) method. 
Each method will be explained separately in Sec.~III.

In Sec.~IV, 
we compare the results together with future directions for improvement and we also discuss some issues for the application to the real data. The mock data we used in this article are available from our webpage~\cite{mockWeb}.


\section{Building mock data}
\subsection{Quasi-Normal Modes}\label{section2.1}
The waveform of the ringdown gravitational waves emitted from an excited BH is modeled as 
\begin{equation}
h(t) = A e^{-(t-t_0)/\tau} \cos(2 \pi f_{\rm R} (t-t_0) - \phi_0) \,, 
\label{ringdownbasic}
\end{equation}
where $f_{\rm R}$ is the oscillation frequency, and $\tau$ is the damping time, and 
$t_0$ and $\phi_0$ are the initial time and its phase, respectively. 
The waveform, Eq.~\eqref{ringdownbasic}, is then written as 
\begin{equation}
h(t)= {\Re} [ A e^{-2\pi \mathrm{i} f_{\rm {qnm}} (t-t_0)} ] \,, 
\label{hoft_frfi}
\end{equation}
where we call $f_{\rm {qnm}}= f_{\rm R}-\mathrm{i}f_I$ the quasi-normal mode (QNM, or ringdown) frequency 
($f_{\rm I}>0$ means decaying mode). 
 (Nakano et al.~\cite{NakanoTanakaNakamura2015} uses different signature on $f_I$.)
The parameter $\tau$ is also expressed using a 
quality factor $Q$, or $f_{\rm I}$
\begin{equation}
Q = \pi {f_{\rm R}} \tau, \quad \mbox{or} \quad f_{\rm I} = \frac{1}{2\pi \tau}=\frac{f_{\rm R}}{2Q} \,. 
\end{equation}

In GR, the set of $(f_{\rm R}, f_{\rm I})$ is determined by the (remnant BH) mass, $M_{\rm rem}$, and angular momentum, $M_{\rm rem}^2 \chi$, of the black hole. 
QNM is obtained from the perturbation analysis of BHs, and its fitting formulas are given by~\cite{BertiCardosoWill2006}
\begin{eqnarray}
f_{\rm R}&=&\frac{1}{2\pi M_{\rm rem}}\{ f_1+f_2(1-\chi)^{f_3} \} \,,
\label{fitting1}\\
Q&=&\frac{f_{\rm R}}{2f_{\rm I}}=q_1+q_2(1-\chi)^{q_3} \,,
\label{fitting2}
\end{eqnarray}
where $f_i$ and $q_i$ are the fitting coefficients. 
For the most fundamental mode, which is of the spherical harmonic index $\ell=2$, $m=2$, the
fitting parameters are $f_1=1.5251$, $f_2=-1.1568$, $f_3=0.1292$, $q_1=0.7000$, $q_2=1.4187,$ and $q_3=-0.4990$.

If we recover the units, 
\begin{equation}
f_{\rm R} (M,\chi)\mbox{~[Hz]}=\frac{c^3}{2\pi G M_{\rm rem}}\{ f_1+f_2(1-\chi)^{f_3} \} \,,
\label{fitting1b}
\end{equation}
where $c$ and $G$ 
are the speed of light and the gravitational constant, respectively.
From this equation, at linear order, the uncertainties in $f_{\rm R}$ and $Q$ are related
to those in mass and angular momentum as 
\begin{equation}
\frac{\Delta f_{\rm R}}{f_{\rm R}}=-\frac{\Delta M_{\rm rem}}{M_{\rm rem}}+\frac{f_2f_3(1-\chi)^{f_3-1}}{f_1+f_2(1-\chi)^{f_3}}\Delta \chi \,.
\end{equation}
Similarly, from Eq.~\eqref{fitting2}, we get
\begin{eqnarray}
{\Delta Q}&=&\left(\frac{\Delta f_{\rm R}}{f_{\rm R}}-\frac{\Delta f_{\rm I}}{f_{\rm I}}\right) Q \\
&=& q_2 q_3(1-\chi)^{q_3-1} \Delta \chi \,.
\end{eqnarray}

In modified gravity theories 
the final fate of binary mergers may not be a black hole. 
There are various possibilities of modification of gravity. The most generic test for the 
deviation from general relativity would be just checking the consistency between the 
observed data and the predicted waveform based on general relativity. 
However, such a generic test will not be very sensitive. 
If we focus on some class of modification, we would be able to perform a much better test. 
Here, we assume that even if the gravity is modified the ringdown waveform is characterized by the set of $(f_{\rm R}, f_{\rm I})$. 
For the same inspiral/merger waveform, which predicts the formation of 
a black hole with $M$ and $\chi$ in GR, however, the values of $(f_{\rm R}, f_{\rm I})$ may be different. 
Under this assumption, one can test GR by comparing $(f_{\rm R}, f_{\rm I})$ predicted from the data in the inspiral/merger phase in GR 
with those directly extracted from the data in the ringdown phase.
For this purpose, we wish to minimize the error in the determination of $(f_{\rm R}, f_{\rm I})$ from the data in the ringdown phase 
independently of the information contained in the inspiral/merger phase.

\subsection{Mock data}\label{section2.2}
The fundamental question we raise here is whether or not one can detect the deviation 
from the GR prediction, in case only the ringdown frequency is modified. 
If the breakdown of GR occurs only in the extremely strong gravity regime 
such as the region close to the BH event horizon, modification of gravity might be 
completely irrelevant to the evolution during the inspiral/merger phases. 
Even in such cases, the deviation from the GR prediction may show up in 
the ringdown waveform. This gives a good motivation to develop a method to identify the ringdown 
frequency without 
referring to the information from the inspiral/merger phases.

There are many proposals for the method to extract the ringdown frequency  and its damping time scale.
In order to compare the performance of various methods by a blind test, we construct some test data which have artificially 
modified ringdown frequency. 

We adopt the following strategy for preparing the data. 
We take the inspiral/merger waveform from SXS Gravitational Waveform 
Database~\cite{Mroue:2013xna,SXSwaveforms}
(there are also available catalogs for BBH GWs in Refs.~\cite{Jani:2016wkt,GATECHwaveforms,Healy:2017psd,RITwaveforms}), 
and the ringdown waveform modified from GR case is merged. 
Then, noise is added so as to reproduce the ideal LIGO noise curve (with the signal-to-noise ratio (SNR) $=$ 20, 30 or 60).
In doing so, we focus on the fact that the time evolution of the amplitude and the frequency 
of the GR waveform is rather smooth if the spin precession can be neglected. 
Our basic assumption is that this smoothness is maintained 
even if we consider modification of the complex QNM frequencies. 
Then, the modified waveform cannot have a large variety. 

We define a normalized time coordinate $x=(t-t_p)/M$, 
where $t_p$ denotes the time that the GW amplitude
has its peak, and just modify the GW strain after the peak time.
This is a reasonable assumption because, if the inspiral/merger 
parts are also modified, we can detect the deviation from GR even in case when 
we cannot extract the QNM frequency from the gravitational wave data. 
Note that $M$ is the initial total mass of the binary,
and we will specify it later to generate the mock data set.
In the following, we take the simplification of considering 
only the $(\ell=2,\,m=2)$ GW mode.

To create the mock data, 
the total mass $M$ is randomly selected from the 
range between $50M_\odot$ -- $70M_\odot$ with uniform probability. 
The parameters characterizing the ringdown waveform, $f_{\rm R}$ and and $f_{\rm I}$, are modified from the GR value within $\pm 30\%$
and $\pm 50\%$, respectively. Uniform probability distribution is assumed for both parameters. 
We present two independent ways to generate the mock data below. 

\subsubsection{Set A}
In the case of set A, the modification is strictly limited to the time 
domain after the peak of the GW amplitude. 
First, as for the GW amplitude $A_{22}=r|h_{22}|/M$,
we introduce the following fitting function
\begin{eqnarray}
\label{amplitude:nakano}
A_{22}(t) &=&
\frac{ A^{\rm GR}_{22}(t_p) +a_0\,x + a_1\,{x}^{2} }
{ 1 - \left( M\,\omega_{\rm I}- a_0/A^{\rm GR}_{22}(t_p) \right) x + a_2\,{x}^{2} }
\cr && \times
\exp(-M\,\omega_{\rm I}\,x) \,,
\end{eqnarray}
where we have three fitting parameters $a_0, a_1$ and $a_2$, 
which are chosen to reproduce the amplitude of the SXS waveform $A^{\rm GR}_{22}(t)$ 
when the adjustable parameter $\omega_{\rm I}=f_{\rm I}/(2\pi)$ 
is set to the GR value $\omega^{\rm GR}_{\rm I}$ calculated from the remnant BH mass and spin
by using Refs.~\cite{BertiCardosoWill2006,BertiQNM}
(In the following, the QNM frequency will often be written using angular frequency $\omega = 2\pi f$). 
For example, for SXS:BBH:0174~\cite{SXSwaveforms}, we have 
\begin{eqnarray}
&& a_0 = 0.0183650 \,, \quad a_1 = 0.000998244 \,, 
\cr &&
a_2 = 0.00184509 \,,
\end{eqnarray}
with $A^{\rm GR}_{22}(t_p)=0.286987$ and 
$M \omega^{\rm GR}_{\rm I} = 0.0815196$. 
The above fitting function is chosen as such that
the first derivative of the amplitude is zero at the peak ($x=0$).
By changing $\omega_{\rm I}$, we can create a mock data. 

Second, as for the GW frequency $\omega_{22}(t)$, which is defined by 
the time derivative of the GW phase and supposed to be positive, 
we use the fitting function
\begin{eqnarray}
\label{frequency:nakano}
M \omega_{22}(t) 
&=& \left( M \omega^{\rm GR}_{22}(t_p) - M \omega_{\rm R}
+b_0\,x + b_1\,{x}^{2} + b_2\,{x}^{3} \right)
\cr && \times 
\exp \left[ {\frac { \left( b_0 + b_3 \right) 
x}{M (\omega_{\rm R} - \omega_{22}(t_p))}} \right]
\cr &&
+M \omega_{\rm R} \,, 
\end{eqnarray}
where $\omega^{\rm GR}(t)$ is the frequency extracted from the phase of 
the numerically determined GR template. 
With this fitting function, the smoothness of GW phase is $C^2$ at the peak time.
The three fitting parameters are, again for SXS:BBH:0174, 
\begin{eqnarray}
&& b_0 = -0.0507805 \,, \quad b_1 = -0.00276104 \,,
\cr
&& b_2 = -0.000479913 \,, \quad b_3 = -0.00492361 \,,
\end{eqnarray}
with $M\omega^{\rm GR}_{22}(t_p)=0.375598$.
The mock data is created by changing the input
$\omega_{\rm R}=f_{\rm R}/(2\pi)$ from the GR value calculated from
the remnant BH mass and spin, {\em e.g.}, $M \omega^{\rm GR}_{\rm R}=0.582652$ 
for SXS:BBH:0174.

\subsubsection{Set B}
The modification of the second set is not strictly restricted to the 
time period after the peak of the amplitude. 
As another smooth interpolation, we adopt
the following modified amplitude 
\begin{equation}
A_{22}(t)=\frac{A^{\rm GR}_{22}(t)}{1+e^{4M\omega_{\rm I}^{\rm GR}x}}
+\frac{A^{\rm RD}_{22}(t)}{1+e^{-4M\omega_{\rm I}^{\rm GR}x}}\,,
\label{amplitude:tanaka}
\end{equation}
with 
\begin{equation}
A^{\rm RD}_{22}(t)=
\frac{{\cal A}_{22}}{1+e^{-M\omega_{\rm I}^{\rm GR}x}+e^{M\omega_{\rm I}\,x}}\,,
\end{equation}
and the overall amplitude ${\cal A}_{22}$  
determined so that the GR case fits well.  

The frequency is also given in a similar simple manner by 
\begin{equation}
\omega_{22}(t)=\frac{\omega^{\rm GR}_{22}(t)}{1+e^{4M\omega_{\rm I}^{\rm GR}x}}
+\frac{\omega^{\rm RD}_{22}(t)}{1+e^{-4M\omega_{\rm I}^{\rm GR}x}}\,,
\label{frequency:tanaka}
\end{equation}
with the GR frequency $\omega^{\rm GR}_{22}(t)$ and 
\begin{equation}
\omega^{\rm RD}_{22}(t)=\omega^{\rm GR}(t_p)+
\frac{\omega_{\rm R}-\omega^{\rm GR}(t_p)}{1+e^{-2M\omega_{\rm I}\,x}}\,.
\end{equation}
For the generation of the set B mock data, we used SXS:BBH:0002, 0004 and 0007. 

\begin{figure}[!th]
  \begin{center}
    \includegraphics[width=.45\textwidth]{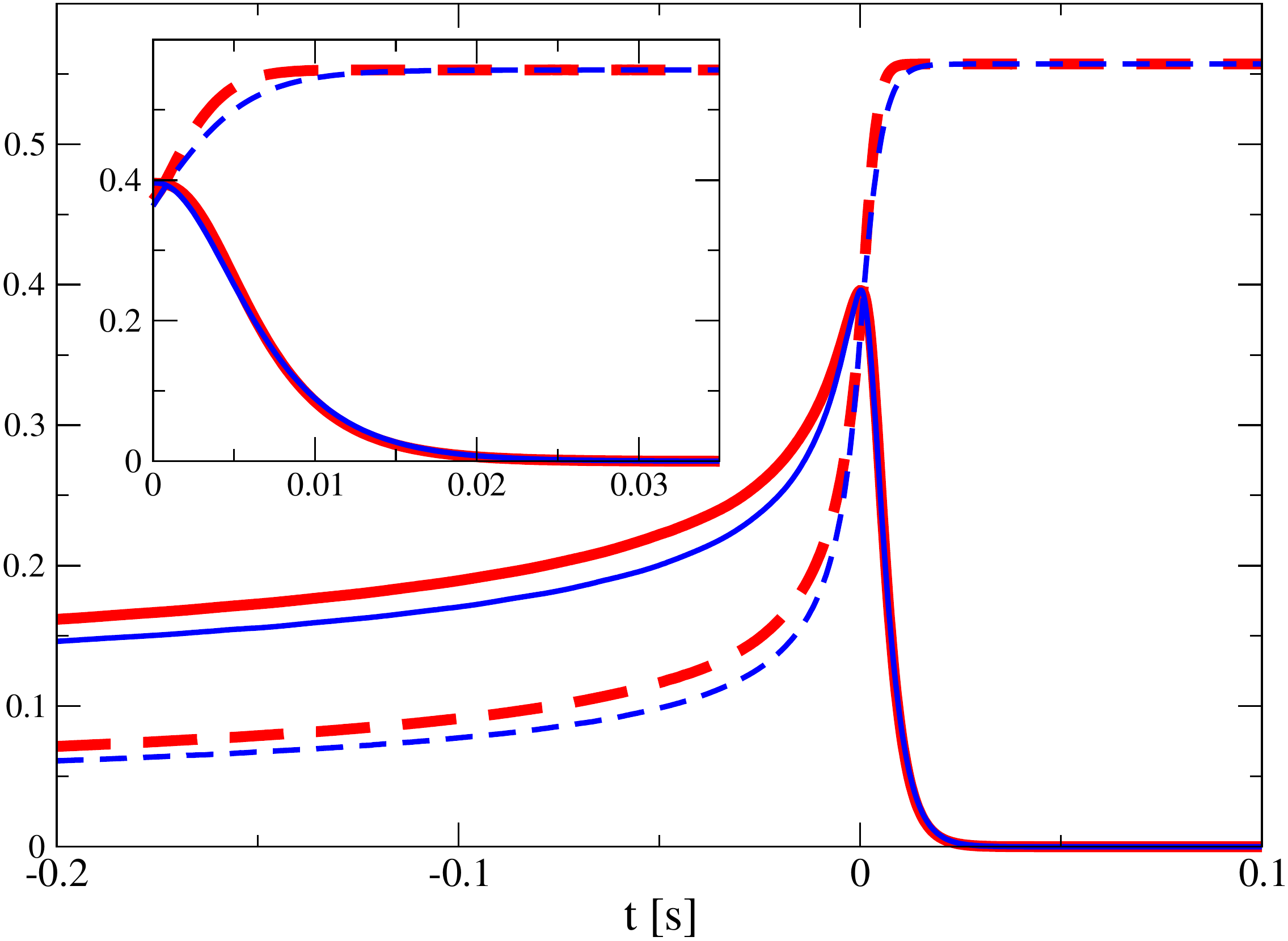}
\caption{Examples of the set A and B.
The inset shows the ringdown part.
Here, a set A ((red) thick) with SXS:BBH:0174 and a set B ((blue) thin) with SXS:BBH:0002 are shown.
The solid lines denote the modified amplitude $A_{22}(t)$,
and the dashed lines are the GW frequency $\omega_{22}(t)/(2\pi)$.
The total mass is $M=60M_{\odot}$,
and the real and imaginary parts of the ringdown frequency are 300\,Hz and 40\,Hz, respectively.
The real frequency is obtained by multiplying by 538.609\,Hz, and the real amplitude of the set A is derived by dividing by 1.37903.
The large difference in the inspiral phase is due to the difference of the binary parameters.}
\label{fig:example}
  \end{center}
\end{figure}

In Fig.~\ref{fig:example}, we show examples of the set A and B with a same ringdown frequency. It is noted that
the binary parameters are different 
between SXS:BBH:0174 (set A) and SXS:BBH:0002 (set B)
in Fig.~\ref{fig:example}.

\section{Various methods for identifying ringdown mode}
\label{section:3}
\subsection{Matched filtering with ringdown part (MF-R)}
We perform the matched filtering analysis using simple damped sinusoidal templates, 
which are given by 
\begin{equation}
\hat h(t) = 
\begin{cases}
& 0 \quad (t<t_0) \\
& \frac{1}{N} e^{-\omega_{\rm I} (t-t_0)} \cos [\omega_{\rm R} (t- t_0) -\phi_0 ] \\
& \hspace{4.7cm} (t \ge t_0) \,,
\end{cases}
\label{template}
\end{equation}
where $t_0$ and $\phi_0$ are the starting time and the initial phase of the template, respectively.
The normalization constant $N$ is chosen so as to satisfy $(\hat h| \hat h) =1$, where 
the inner product is defined by 
\begin{equation}
    (h_1|h_2)=2\int_0^\infty \frac{\tilde h^*_1(f) \tilde h_2(f)+\tilde h_1(f) \tilde h^*_2(f)}
     {S_n(f)} df\,.
     \label{innerproduct}
\end{equation}
Here, $S_n(f)$ is the noise spectral density and the Fourier transform of $h(t)$ is defined by 
$\tilde h(f)=\int dt\, e^{2\pi i f t}h(t)$,
and $*$ denotes the complex conjugate. 
We consider to maximize the inner product between the GW data $s(t)$, which contains the signal and the noise, 
and the template $h(t)$ over the parameters $\omega_{\rm R}$, $\omega_{\rm I}$, $t_0$ and $\phi_0$. 
The SNR against the initial phase of the template can be maximized by rewriting the template in the following form,
\begin{equation}
h(t) =\frac{1}{N} \left (h_c \cos\phi_0 + h_s \sin \phi_0 \right ) \,,
\end{equation}
where
\begin{equation}
\begin{array}{cc}
h_c &= e^{-\omega_{\rm I} (t-t_0)} \cos [\omega_{\rm R}(t-t_0)] \,,\\
h_s & =e^{-\omega_{\rm I} (t-t_0)} \sin [\omega_{\rm R}(t-t_0)] \,.
\end{array}
\end{equation}
Then, the maximum of the SNR against the initial phase $\phi_0$ can be given as~\cite{Nakanoetal2004},
\begin{equation}
\rho^2|_{\mbox{max} \phi_0}= \frac {(s|\hat{h}_c)^2+(s| \hat{h}_s) ^2 -2 (s| \hat{h}_c)(s| \hat{h}_s)(\hat{h}_c|\hat{h} _s)}{1-(\hat{h}_c|\hat{h}_s)^2} \,,
\label{rho}
\end{equation}
where
\begin{equation}
\hat{h}_c=\frac{h_c}{\sqrt{(h_c|h_c) }} \,, \quad \hat{h}_s=\frac{h_s}{\sqrt{(h_s|h_s) }} \,.
\end{equation}
The phase $\phi_0$ that gives the maximum $\rho$ is given by
\begin{equation}
\tan \phi_0 = \frac{(h_c|h_s)(h|h_c)-(h_c|h_c)(h|h_s)}{(h_c|h_s)(h|h_s)-(h_s|h_s)(h|h_c)} \,.
\end{equation}
Then, we are left with three parameters to explore. 
Since the best choice of the starting time of QNM is unknown,
we vary $t_0$ from the merger time $t_c$ to $t_c+$ something.
Then, we search for the best fit values of the parameters $(f, Q)$ for each $t_0$.  
Finally, we calculate median values of $\{f(t_0),Q(t_0)\}$ 
which we regard as our estimate of the QNM frequency.

\subsection{Matched filtering with both merger and ringdown parts (MF-MR)}
\label{section:3.2}
The plain matched filtering using the damped sinusoidal waveform, which was introduced in the preceding 
subsection, has the difficulty in choosing the appropriate starting time, $t_0$. On one hand, 
if $t_0$ is chosen to be too early, we pick up lower frequency oscillations before the QNM starts to dominate. 
On the other hand, if $t_0$ is chosen to be too late, the signal has already become very faint and is buried in 
noise. Therefore, it is likely that this method is not the optimal method to determine the QNM frequency 
from the data.  

The basic idea for the improvement of matched filtering is as discussed in Ref.~\cite{Nagar:2016iwa}.
If we know the modified ringdown waveform in advance, we can construct the 
best linear filter that produces the largest SNR by using it. 
As in the construction of our mock data in Sec.~\ref{section2.1}, 
here we also assume that this smoothness is maintained 
even for the modified waveform. Then, the variety of the possible waveforms 
would be effectively limited well. 

To obtain a better fit, one may think it would be necessary to introduce, at least, 
two more parameters in addition to $(\omega_{\rm R}, \omega_{\rm I})$, i.e., the 
amplitude of QNM relative to that of the inspiral/merger phases, 
and the transition rapidity to reach the final QNM frequency. 
It might be reasonable to perform the matched filtering using 
this generalized waveform including the inspiral/merger phases. 
To find the best fit parameter values in the four parameter space is doable. 
However, we adopt the simplifications of neglecting these additional parameters here, 
in order to reduce the computational cost in the present analysis, leaving this 
possible extension to our future work. 
To reduce the impact of neglecting the relative amplitude of the QNM, 
we make use of the fact that the inspiral/merger parts are basically unchanged 
for all modified templates. 
Namely, we introduce the following sharp window function 
\begin{equation}
    W(t)=\frac{1}{1+e^{-50(t-t_c)\omega_{\rm I}^{\rm GR}}}\,,   
\end{equation}
to make the relative amplitude between the inspiral-merger phase and the ringdown phase 
almost irrelevant, 
instead of introducing one additional model parameter. 

The procedure is summarized as follows. 
We first calculate the whitened signal and template multiplied by the window function, 
which are more explicitly defined by 
\begin{equation}
    \hat s(t)=W(t)\int df e^{-2\pi i ft} \frac{\tilde s(f)}{\sqrt{S_n(f)}} \,,
\end{equation}
and
\begin{equation}
    \hat h(t)= N W(t)\int df e^{-2\pi i ft} \frac{\tilde h(f)}{\sqrt{S_n(f)}} \,, 
\end{equation}
where $N$ is the normalization factor that is determined so as to satisfy 
$(\hat h,\hat h)^{(w)}=1$, where $(*,*)^{(w)}$ is the inner product in Eq.~\eqref{innerproduct}
with $S_n(f)$ replaced with unity. 
After this preprocessing, the correlations between the data and templates are calculated as in the 
case of the standard matched filtering, besides
the point that the inner product  $(*,*)^{(w)}$ is used instead of  $(*,*)$. 
Here, we simply choose the phase such that maximize the 
signal to noise ratio $\rho$ for each template, 
instead of marginalizing over these parameters. 
The origin of the time coordinate is not varied. 
To obtain the distribution of the parameters $\omega_{\rm R}$ and $\omega_{\rm I}$, we 
simply used the probability given by $\propto \exp(\rho^2/2)$, which corresponds 
to the posterior distribution for the flat prior ansatz. 
\footnote{The Likelihood function would be defined by the probability of the 
model parameters $\bm{\theta}$ when a data $D$ is provided, $P(\bm{\theta}|D)$. 
Using the Bayes' theorem, we can write $P(\bm{\theta}|D)$ as 
$P(\bm{\theta}|D)=\frac{P(D|\bm{\theta})P(\bm{\theta})}{P(D)}. $
When we assume a flat prior distribution, i.e., $P(\bm{\theta})=$constant, 
we have  $P(\bm{\theta}|D)\propto P(D|\bm{\theta})$. 
The probability that a data $D$ is described by a normalized template 
$\hat h(\bm{\theta})$ and the noise $n$ as $\rho \hat h+n$ would be given by 
$P(D|\rho,\bm{\theta})\propto \exp[-(n|n)/2]=\exp[-(D-\rho \hat h(\bm{\theta})|D-\rho \hat h(\bm{\theta}))/2]$. 
Marginalizing this probability with respect to $\rho$, we find 
$P(D|\bm{\theta})\propto \exp[(D|\hat h(\bm{\theta}))^2/2]$. 
When we have some unfocused extra parameters, we need to marginalize 
the probability over these extra parameters. Here the coalescence time, 
the overall phase are such parameters. For the coalescence time, we just 
adopt the maximum value, in place of the marginalization, which we expect 
will not cause any significant error.}

To perform the correlation analysis, we also need to specify a template waveform 
which includes the real and imaginary part of the QNM frequency as free parameters. 
To obtain the necessary template waveform, we shall use the same prescription as 
set B that is used to generate a half of the mock data. We understand that this makes 
the comparison for set B unfair, but 
one purpose of testing the improved matched filtering method arranged in this manner 
is to give a relevant standard to evaluate the efficiency of the other methods. 
The standard matched filtering using the damped sinusoidal wave as templates
might be too naive to use it as the standard reference to assess the performance of the other methods. 
This improved matched filtering method is actually guaranteed to give the best linear filtering. 
Therefore, we think that the results obtained by this method offers a good reference to evaluate 
the performance of the other methods. 

\subsection{Hilbert-Huang transformation (HHT) method}
\newcommand{\argmin}{\mathop{\mathrm{argmin}}\limits}
\newcommand{\argmax}{\mathop{\mathrm{argmax}}\limits}
\subsubsection{Basic idea}
The Hilbert-Huang transform (HHT) is a time-frequency analysis method~\cite{HuangPRSL1998},
which is constructed under the aim to manipulate non-stationary and/or non-linear system.
Some applications of the HHT to the data analysis of gravitational waves have been proposed
~\cite{Jordan2007, Alex2009, TakahashiAADA2013, KaneyamaPRD2016, SakaiPRD2017}.
The HHT is based on a signal analysis by the Hilbert transform.
We describe the concept of the signal analysis by the Hilbert transform and its difficulty to be applied to real-world signals,
  and then explain how the HHT overcomes the difficulty.

Letting $\check{s}(t)$ be the Hilbert transform of a signal $s(t)$,
  it is defined by
\begin{align}
    \check{s}(t) = \frac{1}{\pi}\, \mathrm{PV} \int \mathrm{d}t' \, \frac{s(t')}{t - t'} \,,
\end{align}
where PV denotes the Cauchy principal value.
The complex signal $z(t)$,
 which is defined by $z(t) = s(t) + \mathrm{i} \check{s}(t)$,
 can be represented by the exponential form:
\begin{align}
    z(t) = a(t) \mathrm{e}^{\mathrm{i} \phi(t)} \,,
\end{align}
where $a(t)$ and $\phi(t)$ are defined by
\begin{align}
    a(t) &= \sqrt{s(t)^2 + \check{s}(t)^2} \,,
    \\
    \phi(t) &= \arctan \left( \frac{\check{s}(t)}{s(t)} \right) \,.
\end{align}
Therefore,
\begin{align}
    s(t) = a(t) \cos \phi(t)
\end{align}
 is established.
Only when the signal $s(t)$ is monochromatic over short periods of time,
 $z(t)$ is an analytic signal of $s(t)$,
 in other words,
  the Fourier components of $z(t)$ are the same as $s(t)$ in the positive frequency range,
  and zero in the negative frequency range~\cite{Cohen},
 and then $a(t)$ and $\phi(t)$ are called instantaneous amplitude (IA) and instantaneous phase (IP) of $s(t)$, respectively.
The monochromaticity of $s(t)$ over short periods of time means
 that $a(t)$ has only lower frequency components than $\cos \phi(t)$,
 or $a(t)$ and $\cos \phi(t)$ are the modulator and the carrier of the signal $s(t)$, respectively.
In that case,
 the local mean $m(t)$ of $s(t)$, which is defined by
\begin{align}
    m(t) = \frac{u(t) + l(t)}{2} \,,
\end{align}
where $u(t)$ and $l(t)$ are the upper and lower envelopes of $s(t)$, respectively,
 is zero at any point.
We call this feature \textit{zero-mean}.
An instantaneous frequency (IF) of $s(t)$ is defined by
\begin{align}
    f(t) = \frac{1}{2 \pi} \frac{\mathrm{d} \phi(t)}{\mathrm{d} t} \,.
\end{align}
This analysis to estimate the IA and IF from a signal
 is called the Hilbert spectral analysis (HSA).
The HSA has an advantage of higher resolution than the other time-frequency analysis
  such as the short-time Fourier transform and the Wavelet transform.
However,
 it cannot be applied to most real-world signals,
 because they are basically composites of some components and are not monochromatic.
Huang \textit{et al.}~\cite{HuangPRSL1998} overcame the difficulty by combining a mode decomposition part with the HSA,
 and the method of combination of them is the HHT.

Huang \textit{et al.} developed a method to decompose an input data $x(t)$
 into zero-mean components and a non-oscillatory component.
They named the method empirical mode decomposition (EMD)
 and also named the decomposed zero-mean components intrinsic mode functions (IMFs) of the input data.
Algorithm~\ref{algr:EMD} shows the procedure of the EMD,
 where $c_i(t)$ and $r(t)$ are the $i$th IMF and a non-oscillatory component of $x(t)$, respectively.
The first step is forming the upper envelope $u_{i,j}(t)$ and the lower envelope $l_{i,j}(t)$,
 connecting the maxima and the minima of the data by cubic splines.
Then, the mean $m_{i,j}(t)$ of these envelope
 is subtracted from the input data to obtain the residual $x_{i, (j+1)}$.
When the mean $m_{i,j}(t)$ becomes approximately zero after several iterations,
 $x_{i, j}$ is adopted as the IMF $c_{i}(t)$,
 since it can be considered to be zero-mean.
This criteria $\epsilon_\mathrm{e}$ is a parameter of the EMD.
After all oscillatory components are extracted,
 the residual $r(t)$ is a non-oscillatory component of $x(t)$.
Letting $N_\mathrm{IMF}$ be the number of IMFs of $x(t)$,
 $x(t)$ is recovered by
\begin{align}
    x(t)
        &= \sum_{n=1}^{N_\mathrm{IMF}} c_{n}(t) + r(t) \,.
\end{align}
IMFs, $c_1(t),\, c_2(t),\, \dots,\, c_{N_\mathrm{IMF}}(t)$, are
 in order from the highest to the lowest frequency components.
After the above decomposition,
 the IA and IP of each IMF can be estimated by the HSA.
Consequently,
 letting $a_n(t)$ and $\phi_n(t)$ be the IA and IP of $n$th IMF,
 the data can be expressed as
\begin{align}
    x(t)
        &= \sum_{n=1}^{N_\mathrm{IMF}} a_{n}(t) \cos \phi_{n}(t) + r(t) \,.
    \label{eq:HHT}
\end{align}


\begin{algorithm}
  \caption{Empirical mode decomposition}
  \label{algr:EMD}
  \algsetup{indent=2em}
  \begin{algorithmic}[1]
    \setlength{\itemsep}{3pt}
    \STATE $i=1$, $x_1(t) = x(t)$.
    \WHILE{$x_i(t)$ contains oscillatory components}
      \STATE $j=1$, $x_{i,1}(t) = x_{i}(t)$
      \WHILE{the local mean of $x_{i,j}(t)$ is not zero}
        \STATE $u_{i,j}(t) = (\text{the upper envelope of $x_{i,j}(t)$})$.
        \STATE $l_{i,j}(t) = (\text{the lower envelope of $x_{i,j}(t)$})$.
        \STATE $m_{i,j}(t) = (u_{i,j}(t) + l_{i,j}(t))/2$.
        \STATE $x_{i,(j+1)}(t) = x_{i,j}(t) - m_{i,j}(t)$.
        \STATE $j = j+1$
      \ENDWHILE
      \STATE $c_i(t) = x_{i,j}(t)$
      \STATE $x_{i+1}(t) = x_{i}(t) - c_{i}(t)$.
      \STATE $i = i + 1$.
    \ENDWHILE
    \STATE $r(t) = x_{i}(t)$
  \end{algorithmic}
\end{algorithm}
In this study,
  we used the ensemble EMD (EEMD) as the mode decomposition method.
In the beginning of the EEMD,
 $N_\mathrm{e}$ white noises $\{w^{(m)}(t)\}$ with the standard deviation being $\sigma_\mathrm{e}$ are created,
 and then the IMFs of the noise-added data $x^{(m)}(t) = x(t) + w^{(m)}(t)$ are calculated by the EMD:
\begin{align}
    x^{(m)}(t)
      &= x(t) + w^{(m)}(t) \\
      &= \sum_{n=1}^{N_\mathrm{IMF}} c^{(m)}_{n}(t) + r^{(m)}.
\end{align}
The IMFs of an input data $x(t)$
  are estimated as the mean of the corresponding IMFs of $\{x^{(m)}(t)\}$:
\begin{align}
    c_n(t) = \frac{1}{N_\mathrm{e}} \sum_{m=1}^{N_\mathrm{e}} c^{(m)}_{n}(t).
\end{align}
The EEMD has two parameters $(\sigma_\mathrm{e}, \epsilon_\mathrm{e})$.
$\sigma_\mathrm{e}$ is a standard deviation of added white noise,
  and $\epsilon_\mathrm{e}$ is a convergence-condition constant.
The details of EEMD is shown in Refs.~\cite{Wu2009, KaneyamaPRD2016}.

The basic concept of HHT for the QNM is as follows.
If the QNM is perfectly extracted in the $j$th IMF,
  the IA and IP of the IMF must be expressed by 
\begin{align}
    a_j(t)
        &= A \mathrm{e}^{-(t - t_0)/\tau} \,, \label{eq:QNM_IA}\\
    \phi_j(t)
        &= 2 \pi f_\mathrm{R}(t - t_0) + \phi_0 \,. \label{eq:QNM_IF}
\end{align}
Therefore, we can estimate the QNM frequency by fitting the IA and IP individually.

In reality, the IMF also contains other modes before the QNM starts, and noise components become dominant 
after the QNM is sufficiently damped. Equations \eqref{eq:QNM_IA} and \eqref{eq:QNM_IF} 
do not hold in the merger phase and the noise-dominant segment. 
Therefore, to estimate the QNM frequency with Eqs.~\eqref{eq:QNM_IA} and \eqref{eq:QNM_IF},
 we need to estimate the segment where IA and IP most properly fits the equations.
We constructed a method to estimate the segment, named \textit{QNM-dominant segment} (QDS), and the QNM frequency~\cite{SakaiPRD2017}.
In the method,
 a bandpass filter, whose higher cutoff frequency is properly configurated, will be applied as a preprocessing to extract a QNM into the 1st IMF.

\subsubsection{QDS estimation}\label{subsec:QDS_estimation}
Here,
 we briefly describe how to estimate QDS $[\hat{n}_0, \, \hat{n}_0 + \hat{N}]$.
Note that we represent discrete sequences with brackets, such as $t[n]$ and $a_1[n]$.
Assuming the QNM is extracted in the 1st IMF,
 and its merger time $t[n_\mathrm{m}]$ is known,
 we first search for the longest segment $[n_\mathrm{b},\,n_\mathrm{e}]$ after $n_\mathrm{m}$
 where the $a_1[n]$ decreases monotonically.
For every possible subsegments $[n_0, \, n_0 + N]$ of $[n_\mathrm{b},\,n_\mathrm{e}]$,
 where $N_\mathrm{min} \leq N \leq n_\mathrm{e} - n_\mathrm{b}$,
 we calculate root mean squared errors $\mathrm{RMSE}(n_0,N)$ of fitting $\ln a_1[n]$ with $bt[n] + c$:
\begin{align}
    \mathrm{RMSE}(n_0, N)
      = \min_{b,c} \sqrt{\frac{1}{N} \sum_{n=n_{0}}^{n_{0}+N-1} (\ln a_1[n] - bt[n] - c)^2} \,.
\end{align}
We set $N_\mathrm{min}$ to 5, the same configuration with Ref.~\cite{SakaiPRD2017}.
The optimal $n_0$ for each $N$ is determined by
\begin{align}
    \hat{n}_{0}(N) = \argmin_{n_{0}} \mathrm{RMSE}(n_0, N) \,,
\end{align}
and we define $e(N)$ as
\begin{align}
    e(N) = \mathrm{RMSE}(\hat{n}_0(N), N) \,.
\end{align}
The optimal $N$ is determined as the transition point of a slope of $N$--$e(N)$ plot:
\begin{align}
    \hat{N}
      = \argmin_{N} [\mathrm{Err}(N_\mathrm{min}, N) + \mathrm{Err}(N + 1, n_\mathrm{e} - n_\mathrm{b})] \,,
\end{align}
where $\mathrm{Err}(N_1, N_2)$ is an error of the fitting $e(N)$ with $aN + b$:
\begin{align}
    \mathrm{Err}(N_1, N_2) = \min_{a,b} \sqrt{ \frac{\sum_{N=N_{1}}^{N_{2}} (e(N) - aN - b)^2 }{N_2 - N_1} } \,.
\end{align}
Consequently,
  by letting $\hat{n}_{0} = \hat{n}_{0}(\hat{N})$, 
  the QDS $[\hat{n}_0, \hat{n}_{0} + \hat{N}]$ is estimated.
Then,
  the QNM frequency $f_\mathrm{qnm}$ can be estimated by fitting the IA and IP with Eqs.~\eqref{eq:QNM_IA} and  \eqref{eq:QNM_IF} in the QDS.

\subsubsection{Method}\label{subsec:hht_mehod}

Here, we briefly explain the whole method to estimate QNM frequency from observed strain data $h[n]$.
The outline of the method is described in Algorithm~\ref{algr:HHT-QNM}.

First, we have to determine a candidate sets $F$, $E$, $\varSigma$,
  which are sets of a higher cutoff frequency $f_\mathrm{H}$ of a bandpass filter,
  a convergence criteria $\epsilon_\mathrm{e}$ of the EEMD,
  and a standard deviation $\sigma_\mathrm{e}$ of the added-noise in the EEMD, respectively.
In this study,
  we determined the sets as follows:
\begin{align}
    F
        &= \{ 220,\, 225,\, 230, \dots, 500 \} \, \mathrm{Hz},
    \\
    E
        &= \{ 1 \times 10^{-1}, \, 4 \times 10^{-2}, \, 2 \times 10^{-2}, \dots , 1 \times 10^{-3} \} \,.
    \\
    \varSigma
        &= \{ 1 \times 10^{-3}, \, 4 \times 10^{-4}, \, 2 \times 10^{-4}, \dots , 1 \times 10^{-5} \} \,,
\end{align}
and $f_\mathrm{L}$ is set to 20\,Hz.
For each parameter candidates $(f_\mathrm{H}, \epsilon_\mathrm{e}, \sigma_\mathrm{e})$, 
 a series of processing,
  including a bandpass filter with cutoff frequency $(f_\mathrm{L}, f_\mathrm{H})$, 
  the HHT and the search of QDS,
 is applied to the input strain data $h[n]$. 
After that,
  the optimal set of the parameters will be selected under an objective function $O$.
We used the slope of the linear function obtained by fitting $f_1[n]$ in the range of the searched QDS as the objective function $O$,
 since the IF must be flat in the QDS if a QNM is properly extracted.

\begin{algorithm}
  \caption{Estimation method of the QNM frequency with the HHT}
  \label{algr:HHT-QNM}
  \algsetup{indent=2em}
  \begin{algorithmic}[1]
    \setlength{\itemsep}{3pt}
    \REQUIRE Strain $h[n]$ contains a BBH signal and merger time $t_\mathrm{m}$ is known
    \FORALL{$(f_\mathrm{H}, \epsilon_\mathrm{e}, \sigma_\mathrm{e}) \in F \otimes E \otimes \varSigma$}
      \setlength{\itemsep}{3pt}
      \STATE $h[n] \to h_\text{filtered}[n]$: \\
             \quad apply a bandpass filter with cutoff freq.\ $(f_\mathrm{L}, f_\mathrm{H})$
      \STATE $h_\text{filtered}[n] \to a_1[n] , \phi_1[n]$: \\
             \quad apply the HHT with parameters $(\epsilon_\mathrm{e}, \sigma_\mathrm{e})$.
      \STATE $[\hat{n}_0, \hat{n}_0 + \hat{N}]$: \\
             \quad search the QNM dominant segment (QDS) in $a_1[n]$
      \STATE $\hat{f}_\mathrm{qnm}(f_\mathrm{H}, \hat{\epsilon}_\mathrm{e}, \hat{\sigma}_\mathrm{e})$: \\
             \quad estimate the QNM freq.\ by fitting $a_1[n]$ , $\phi_1[n]$ in the QDS
    \ENDFOR
    \STATE $(\hat{f}_\mathrm{H}, \hat{\epsilon}_\mathrm{e}, \hat{\sigma}_\mathrm{e}) = \argmin_{f_\mathrm{H}, \epsilon_\mathrm{e}, \sigma_\mathrm{e}} O(f_\mathrm{H}, \hat{\epsilon}_\mathrm{e}, \hat{\sigma}_\mathrm{e})$: \\
            \quad select the set of the parameters that optimizes an objective function $O$.
    \STATE $\hat{f}_\mathrm{qnm} = \hat{f}_\mathrm{qnm}(\hat{f}_\mathrm{H}, \hat{\epsilon}_\mathrm{e}, \hat{\sigma}_\mathrm{e})$: \\
            \quad the value of the selected combination is the estimated value of this method.
  \end{algorithmic}
\end{algorithm}

\subsection{Autoregressive modeling (AR) method}
\subsubsection{Basic idea}
Autoregressive (AR) method is well-known time-sequence analysis method
which are used in, e.g., acoustic signal processing \cite{BookMarple}.
Suppose we have the signal data of a segment, $x_n = x(n \Delta t)$,  $(n=1,2,\cdots, N)$. 
The main idea is to express the signal $x_n$ with its previous $M (<N)$ data, 
\begin{eqnarray}
x_n 
&=&\sum_{j=1}^M a_j x_{n-j}+\varepsilon \,, 
\label{eq_AR}
\end{eqnarray}
where  
$a_j$ and $M$ are the coefficients and the order of AR model, respectively, and 
$\varepsilon$ is the residual treated as white-noise in this modeling. 
If the data $x_n$ is damped sinusoidal wave without noise, then we analytically can express $x_n$ with $M=2$. Even when the data includes noise, we expect to extract the actual signals by tuning $N$ and $M$. 
There are various methods proposed to determine $a_j$ and $M$. In this article, 
we present the results using Burg method for $a_j$ and 
final prediction error (FPE) method for $M$. 
The details and other trials are in Ref.~\cite{YamamotoShinkai}.

Once the model~\eqref{eq_AR} is fixed, we then reconstruct wave signal from Eq.~\eqref{eq_AR} and analyze it. 
By setting $z(f)=e^{2 \pi i f  \Delta t }$, the power spectrum of the wave signal can be expressed as
\begin{eqnarray}
p(f)
=\sigma^2 \left|1-  \sum_{j=1}^M a_j z^{-j}  \right|^{-2} \,, 
\label{eq_ARfreq}
\end{eqnarray}
where $\sigma$ is the variance of $\varepsilon$. The resolution of frequency in Eq.~\eqref{eq_ARfreq} is not
limited by the length of the original data set, so that AR method is expected to identify signal frequency more precisely 
than the standard (short-) Fourier decomposition.  

From Eq.~\eqref{eq_AR}, 
the (local) maximums of the spectrum, $p(f)$, are given at 
\begin{equation}
F(z) = 1- \sum_{j=1}^M a_j z^{-j} \approx 0. 
\label{eq_chara}
\end{equation}
This is a $M$-th order polynomial equation. 
The solutions of the characteristic equation,  
$F(z)=0$, 
also express the fundamental modes which consist the data segment. 
By interpreting the $M$ solutions as 
$z_k=e^{2 \pi i f_k \Delta t}$ ($k=1,\cdots, M$), 
we get the fundamental frequencies from 
the real part of $f_k$, and damping rates from the imaginary part of $f_k$. (Actually, $|z_k| \leq 1 $ is expected for the expression~\eqref{eq_AR} to be stable.)
Therefore, AR method can determine the frequencies and damping rates of quasi-normal modes from the data themselves.  

\subsubsection{Method}
We divided the given mock data into segments of the length of $\Delta T=1/128$\,sec ($N=32$). 
The neighboring segments are largely overlapping shifted only by 4 points. 
For each segment, we modeled the data with Eq.~\eqref{eq_AR} with Burg and FPE methods. 
Normally $M$ falls into the range 2--5.


We then get the power-spectrum $p(f)$ from Eq.~\eqref{eq_ARfreq} at each segment, and locate its local maximums $f_1, f_2, \cdots$ with their one-sigma widths. 
We also solve Eq.~\eqref{eq_chara} at each segment (which is at most 5-th order polynomial equation), 
and identify the solution $z_k$ of which real part of frequency is the closest to $f_1, f_2, \cdots$. 
 
We list these solutions $z_k$ of each segment, and check whether they remain almost unchanged over several segments.
If the successive segments has a common frequency mode within one-sigma width, then we make a short list
as the candidates for ring-down modes.

We see sometimes a segment is full of noises and shows quite different numbers from neighboring segments. Most cases, however, after the time of black hole merger, we can identify one common frequency which overlaps within one-sigma width for several data segments. 



\subsection{Neural network (NN) method}\label{section:3.5}

\subsubsection{Convolutional Neural Network}

In this challenge, we use a ``convolutional neural network'' (CNN)
which can extract local patterns of the data.
CNNs are often used for the image recognition
and we expect CNNs can be applied to the GW data analysis~\cite{GeorgeHuerta}.
We try various CNNs which have different structure, layers, neurons and filters.
The final configuration of the CNN which is used here is shown in Table~\ref{structureofnn}.

In general, the input and output data of a convolutional layer have multi channel.
The data with multi channel have multi values in each pixel (e.g., RGB components of images).
In a convolutional layer,
convolutions of the data containing $L$ channels and the $L'$ filters $h$,
\begin{equation}
z'_{i, l'} = \sum_{l=1}^L \sum_{p=1}^H z_{i+p, l} h^l_{p, l'} + b_{i, l'} \,,
\end{equation}
are calculated to extract the local patterns.
Here, $z$ and $z'$ are the input and output vectors of the layer, respectively.
The number and the length of the filters, $L'$ and $H$, are fixed before training
and the coefficients $h$ and biases $b$ are optimized in the training procedure.
In this work, we use four convolutional layers.
The lengths of the filters are 32, 16, 8, 8,
and the numbers of the filters are 64, 128, 256, 512, respectively.

A pooling layer, often placed after a convolutional layer,
compresses information, combining a few pixels into one pixel.
In this work, we use the max pooling,
\begin{equation}
z'_i = \max_{k=1,\dots,p} z_{si+k} \,.
\end{equation}
Here, $s$ is the stride, and $p$ is the size of the pooling filter.
We set $s=p=2$.

In most cases, 
dense layers, which are linear transformations,
\begin{equation}
z'_i = \sum_{j=1}^N w_{ij}z_j + b_i \,, 
\end{equation}
with $N$ being the number of input values,
are located after convolutional layers.
The weights $w$ and biases $b$ are optimized in training.

An activation function plays an important role to
carry out nonlinear transformations.
In this work, we use the rectified linear unit (ReLU),
\begin{equation}
z' = h(z) = \max (z, 0) \,,
\end{equation}
as activation functions.

\begin{table}[ht]
\caption{The configuration of the CNN we use.
$(x, y)$ means that the data each layer returns has $x$ points and $y$ channels.
Each input data has only 1 channels, $h_+$, 
which are composed of 512 points.
A "Flatten" layer reshapes a 2-dimensional data to a 1-dimensional data.
}
\begin{tabular}{c|c}
\hline 
layer & dimension \\ \hline
Input & (512, 1) \\
Conv & (481, 64) \\
Pooling & (240, 64) \\
ReLU & (240, 64) \\
Conv & (225, 128) \\
Pooling & (112, 128) \\
ReLU & (112, 128) \\
Conv & (105, 256) \\
Pooling & (52, 256) \\
ReLU & (52, 256) \\
Conv & (45, 512) \\
Pooling & (22, 512) \\
ReLU & (22, 512) \\
Flatten & 16$\times$512 \\
Dense & 256 \\
ReLU & 256 \\
Dense & 2 \\
Output & 2 \\
\hline
\end{tabular}
\label{structureofnn}
\end{table}

For an accurate estimation,  weights and biases in dense layers and filters in the convolutional layers need to be optimized using training data.
In the case of supervised learning,
each training data is a pair of input data and a target vector.
In our work, the input data is the time series of a gravitational wave signal with noise and the target vector is composed of the QNM frequency, $(f_\mathrm{R}^\mathrm{(inj)},\, f_\mathrm{I}^\mathrm{(inj)})$.
For an input, the neural network returns an estimated vector $(f_\mathrm{R}^\mathrm{(pred)},\, f_\mathrm{I}^\mathrm{(pred)})$
and compares it with a target vector.
The loss function is computed to evaluate the error between the estimated and target vectors.
In batch learning,
a group of data, called ``batch" with its size $N_\mathrm{b}$ fixed before training,
is used to define the loss function.
As the loss function, we adopt the mean relative error,
\begin{equation}
J = \frac{1}{N_\mathrm{b}} \sum_{n=1}^{N_\mathrm{b}} \sum_{A \in \mathrm{\{R,I\}}}
\frac{|f_{n,A}^\mathrm{(pred)} - f_{n,A}^\mathrm{(inj)}|}{f_{n,A}^\mathrm{(inj)}}
\end{equation}
We set $N_\mathrm{b}=64$.
As an optimization method, we use the Adam (Adaptive Moment Estimation)~\cite{Adam}.
The hierarchical training, proposed in Ref.~\cite{GeorgeHuerta}, is adopted.
The training starts from using the injected data whose peak SNR
\footnote{The definition of peak SNR is 
$\mathrm{(peak\ SNR)} = \mathrm{(maximum\ amplitude\ of\ template)} / \mathrm{(variance\ of\ noise)}$.}
is 20.0
and gradually decreasing the peak SNR.
At the final stage of training, we use the signal of which peak SNR is ranged from 10.0 to 3.0.

We use the \texttt{PyTorch}~\cite{PyTorch}.
For accelerating learning with NVIDIA GPU, we employ the CUDA deep neural network library (cuDNN)~\cite{cuDNN}.

\subsubsection{Training Dataset}

First, we construct the template bank for the training 
using the modified waveform which is based on the same method as Eqs.~\eqref{amplitude:tanaka} and \eqref{frequency:tanaka}.
For template bank, $f_{\rm R}$ and $f_{\rm I}$ are uniformly placed
in the range 209--378\,Hz and 23--69\,Hz.
The template bank contains 21$\times$21 waveforms.

Next, each waveform is whitened using the aLIGO's design sensitivity (aLIGOZeroDetHighPower).
From each signal, we pick up the segment that consists of 512 points starting from the coalescence time. 
And these whitened waveforms are injected into the white Gaussian noises.
The realization of noises are varied for each training step in order to prevent
that the neural network is overfitted with some particular noise patterns.
Finally, we normalize each segment to have mean 0.0 and variance 1.0.

\section{Comparison and Summary}

\subsubsection{Overview}
We prepare 120 mock data in total by using the method described in Sec.~\ref{section2.2}. A half of them are generated using 
Eqs.~\eqref{amplitude:nakano} and \eqref{frequency:nakano}, and 
the others are generated using Eqs.~\eqref{amplitude:tanaka} and \eqref{frequency:tanaka}. 
We refer to the former as set A and the latter as set B. 
For both sets, we generated 20 mock data, respectively, with overall SNR, 
$\rho_{\rm all} = 60,\, 30$ and $20$. 
The SNR for the ringdown part,  $\rho_{\rm rd}$, 
turned out to be roughly $1/5 \sim 1/3$ of $\rho_{\rm all}$. 
We listed the details of a part of mock data in 
Table~\ref{table_listdata}. 
We calculated $\rho_{\rm rd}$ by the standard inner product for the injected waveform after the peak of the amplitude without noise.

The five challenging groups received 120 data-files of $h(t)$, together with rough information of the merger time, $t_0$, for each data, 
but not with the frequency of the injected ringdown waveform, $(f^{\rm (inj)}_{\rm R},\, f^{\rm (inj)}_{\rm I})$.  The mock data itself is provided with both $+$ mode and $\times$ mode, 
but this time we used only $+$ mode.  
Since $(f^{\rm (inj)}_{\rm R},\, f^{\rm (inj)}_{\rm I})$ are randomly shifted from the values in general relativity, the challengers cannot use the information of inspiral part for their estimation of $(f_{\rm R},\, f_{\rm I})$. 
Some methods (MF-R/MR, AR) can derive the estimation value $(f_{\rm R},\, f_{\rm I})$ with their error bars, while some (HHT, NN) cannot. Therefore we simply compare the results of the estimated (central) values. 

\begin{table}[!ht]
\caption{\label{table_listdata} A partial list of mock data. Set A was generated using Eqs.~\eqref{amplitude:nakano} and \eqref{frequency:nakano}, while set B was from Eqs.~\eqref{amplitude:tanaka} and \eqref{frequency:tanaka}.
The overall SNR $\rho_{\rm all}$, SNR of the ringdown part $\rho_{\rm rd}$, and injected value of the ringdown waveform
$(f^{\rm (inj)}_{\rm R},\, f^{\rm (inj)}_{\rm I})$ are shown. }
\begin{tabular}{p{1cm}p{1cm}p{1cm}p{1cm}p{1cm}}
\hline
data  & \multicolumn{2}{c}{SNR}  & \multicolumn{2}{c}{injected}  \\
  &  $\rho_{\rm all}$ &  $\rho_{\rm rd}$ & $f^{\rm (inj)}_{\rm R}$ & $f^{\rm (inj)}_{\rm I}$\\
\hline
A-01 & 60.0  & 11.87  & 260.68 & 44.58\\
A-02 & 60.0  & 12.82  & 345.16 & 50.49\\
A-03 & 60.0  & 13.31  & 382.53 & 32.58\\
A-04 & 60.0  & 12.49  & 284.18 & 44.73\\
A-05 & 60.0  & 14.25  & 346.20 & 23.07\\
A-06 & 30.0  & ~6.18  & 272.85 & 33.40\\
A-07 & 30.0  & ~6.07  & 272.85 & 44.54\\
A-08 & 30.0  & ~6.05  & 301.89 & 42.24\\
A-09 & 30.0  & ~6.75  & 324.60 & 27.25\\
A-10 & 30.0  & ~6.08  & 282.55 & 37.45\\
A-11 & 20.0  & ~4.59  & 314.24 & 30.58\\
A-12 & 20.0  & ~3.85  & 382.10 & 48.60\\
A-13 & 20.0  & ~4.01  & 249.36 & 47.97\\
A-14 & 20.0  & ~3.98  & 299.32 & 41.88\\
A-15 & 20.0  & ~4.09  & 319.42 & 31.55\\
B-01 & 60.0  & 15.93  & 352.56  & 36.20 \\
B-02 & 60.0  & 15.62  & 210.78  & 42.77 \\
B-03 & 60.0  & 15.31  & 258.83  & 48.42 \\
B-04 & 60.0  & 18.34  & 271.13  & 25.40 \\
B-05 & 60.0  & 15.92  & 291.99  & 34.20 \\
B-06 & 30.0  & ~8.55  & 411.57  & 29.48 \\
B-07 & 30.0  & ~6.78  & 295.78  & 59.38 \\
B-08 & 30.0  & ~7.03  & 312.39  & 59.24 \\
B-09 & 30.0  & ~7.68  & 198.34  & 57.91 \\
B-10 & 30.0  & ~7.81  & 323.32  & 37.86 \\
B-11 & 20.0  & ~5.79  & 208.80  & 39.75 \\
B-12 & 20.0  & ~5.76  & 246.66  & 27.85 \\
B-13 & 20.0  & ~4.46  & 323.71  & 62.51 \\
B-14 & 20.0  & ~5.62  & 215.15  & 33.15 \\
B-15 & 20.0  & ~5.99  & 335.20  & 25.11 \\
\hline
\end{tabular}
\end{table}

\subsubsection{Results and Comparison}
For 120 data, each challenging group handed in $(f_{\rm R}, f_{\rm I})$ as the result of their blind analyses. 
In order to compare five methods, 
we introduce the logarithmic average and variance defined by  
\begin{eqnarray}
&&\overline{\delta\log Q} = \frac{1}{N}\sum_{n=1}^N\left(\log\frac{Q^{\rm (estimate)}_n}{Q^{\rm (inj)}_n}\right)\,,
\cr
&&\sigma(Q) = \left[\frac{1}{N}\sum_{n=1}^N\left(\log\frac{Q^{\rm (estimate)}_n}{Q^{\rm (inj)}_n}\right)^2\right]^{1/2}\,,
\end{eqnarray}
as indicators of the bias and the average magnitude of the 
parameter estimation error, 
where $Q^{\rm (estimate)}_n$ is the estimated value of the quantity $Q$ for the 
$n$th data and $Q^{\rm (inj)}_n$ is the corresponding injected values. 
In Table~\ref{table:logave}, we show  
the values of $\overline{\delta\log f_{\rm R}}$, 
$\sigma(f_{\rm R})$, $\overline{\delta\log f_{\rm I}}$ and  $\sigma(f_{\rm I})$ for the methods we tried. 
We show the results limited to set A on the first law and those limited to set B on the second law. 

We should recall that in the actual implementation of the MF-MR described in Sec.~\ref{section:3.2} we adopt the same modified template that is used to generate the set B mock data. 
Also, the NN method introduced 
in Sec.~\ref{section:3.5} uses the template bank generated in the same way as the set B mock data to train the network. 

The error of MF-R using the simple damped sinusoidal waveform is relatively 
large as expected. In fact, the error of the estimates of $f_{\rm R}$ and $f_{\rm I}$ are the largest among five methods. 
The results of HHT method are not so good, either. 
At least, the current way of using MF-R or HHT for the 
estimate of imaginary part of QNM frequency does not seem 
to be competitive compared with the other methods. 
The performances of the other three methods, 
i.e., MF-MR, AR and NN methods, 
are almost comparable for the imaginary part, while 
the determination of the real part by AR looks better than the other two methods. 
Here we should recall that the comparison with MF-MR and NN is not fair in the case 
of the set B mock data, 
since their base templates are constructed from the set B data. 
The results of MF-MR and NN are better for set B, as expected. 

\begin{figure*}[!th]
  \begin{center}
  \subfigure[~~Real part for Set A]{
    \includegraphics[width=.48\textwidth]{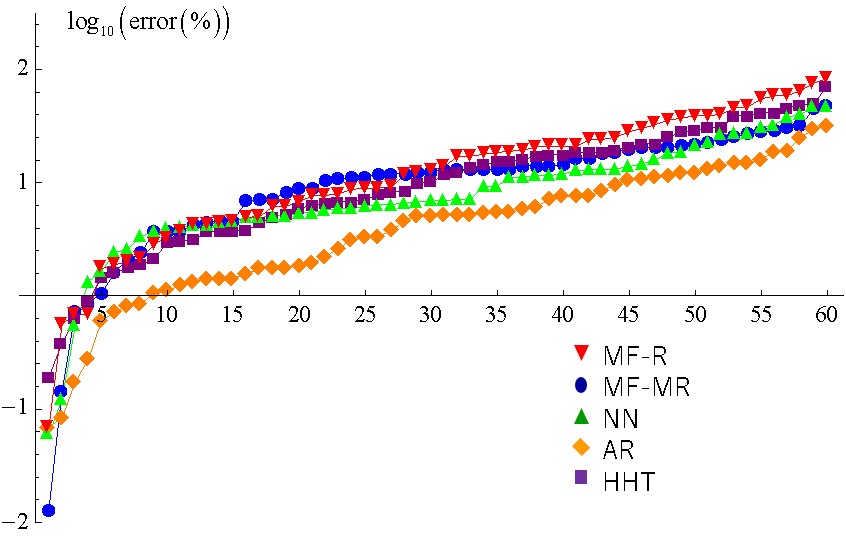}}
  \subfigure[~~Real part for Set B]{
      \includegraphics[width=.48\textwidth]{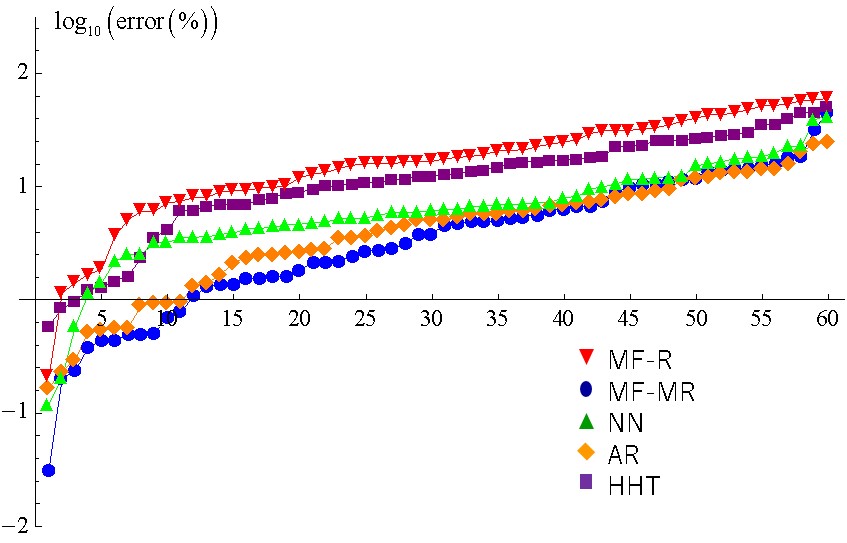}}
   \subfigure[~~Imaginary part for Set A]{
      \includegraphics[width=.48\textwidth]{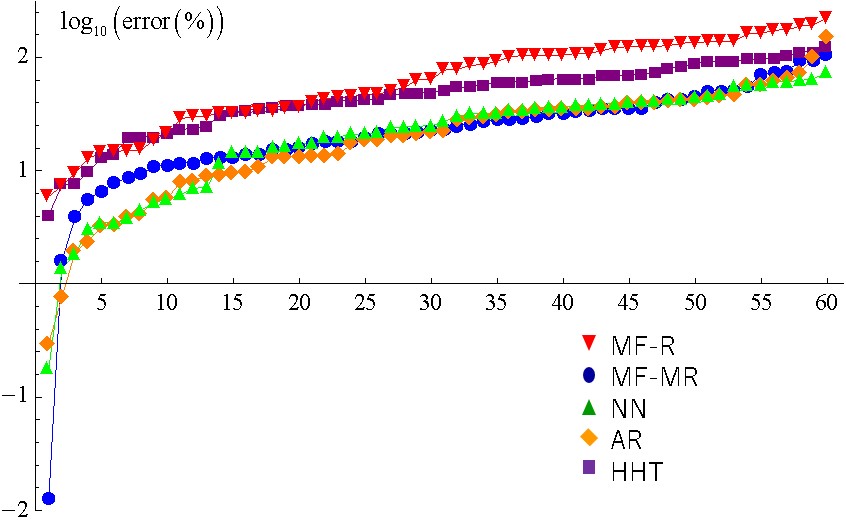}}
   \subfigure[~~Imaginary part for Set B]{
    \includegraphics[width=.48\textwidth]{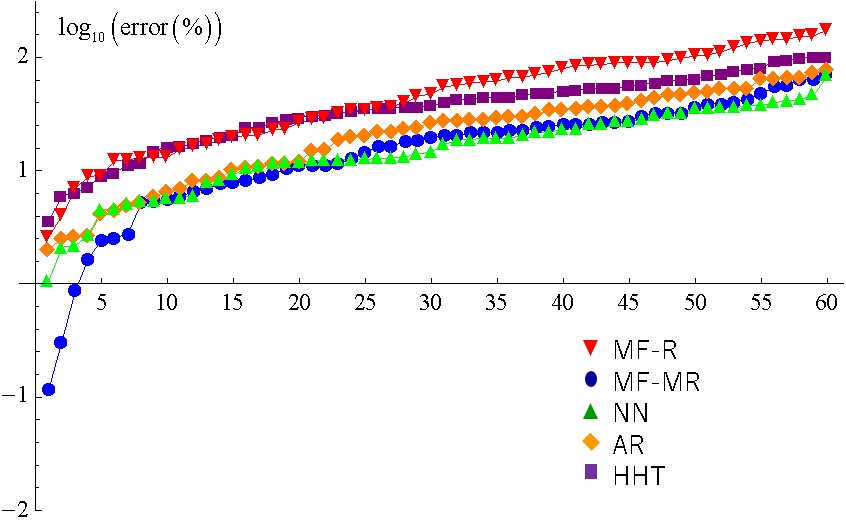}}
\caption{Plots of the base 10 logarithm of the error in the estimate 
for each test data. The data number is 
sorted for each method in the ascending order of the magnitude of the error.
\label{fig:1}}
  \end{center}
\end{figure*}

The variance might be determined by a small number of data with a large error. 
To check if it is the case or not, 
we give plots of the absolute magnitude of the error 
$|\log({Q^{\rm (estimate)}_n}/{Q^{\rm (inj)}_n})|$ 
sorted in the ascending order for each method in Fig.~\ref{fig:1}.
Although the number of data is small, these figures tell that 
the tendencies mentioned above are not the ones that hold only for the data 
with a large error. 

\begin{table}[!t]
\caption{We show  
the values of $\overline{\delta\log f_{\rm R}}$, 
$\sigma(f_{\rm R})$, $\overline{\delta\log f_{\rm I}}$ and  $\sigma(f_{\rm I})$ for various methods. 
The results limited to set A are given on the first law of each method 
while those limited to set B on the second. 
\label{table:logave}}
\begin{center}
\begin{tabular}{lc|cccc} \hline
&& $~\overline{\delta\log f_{\rm R}}(\%)~$ & $~~\sigma(f_{\rm R})(\%)~~$ & $~\overline{\delta\log f_{\rm I}}(\%)~$ & $~~\sigma(f_{\rm I})(\%)~~$\\ \hline\hline   
MF-R &A& -12.88 &  28.36 &  -71.51 &  97.79\\ 
&B&-0.82 &  27.53 &  -46.11 &  75.48 \\ \hline
MF-MR &A& 
6.25 & 17.27 & -12.62 & 37.9
\\ 
&B&2.47 & 10.41 & 7.18 & 27.61
\\ \hline 
HHT &A& -13.38 & 21.91 & -44.11 & 61.58 \\ 
&B& -8.08 & 19.81 & -28.78 & 49.61 \\ \hline 
AR &A& 0.2 & 9.93 & 4.88 & 38.75 \\ 
&B& 1.91 & 8.57 & 6.2 & 34.64\\ \hline 
NN &A& -6.64 & 16.48 & -15.23 & 33.96\\ 
&B&-6.65 & 11.97 & 9.96 & 23.76\\ \hline
\end{tabular}
\end{center}
\end{table}

\begin{figure*}[!th]
  \begin{center}
  \subfigure[~~Both Set A and B]{
    \includegraphics[width=.48\textwidth]{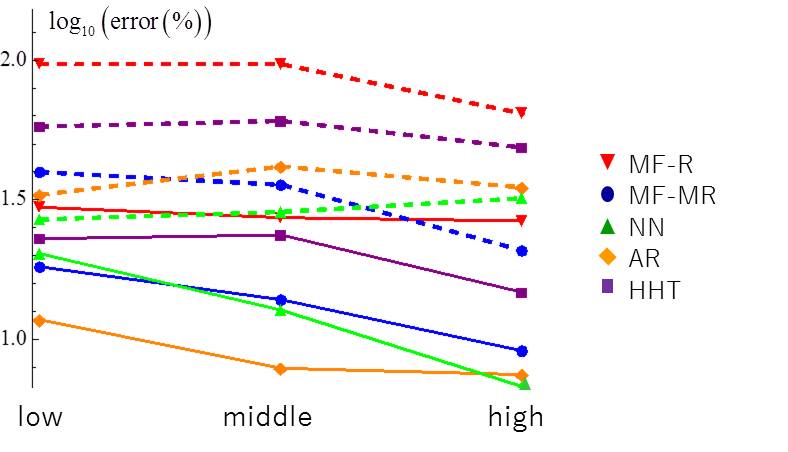}}
  \subfigure[~~Set A only]{
      \includegraphics[width=.48\textwidth]{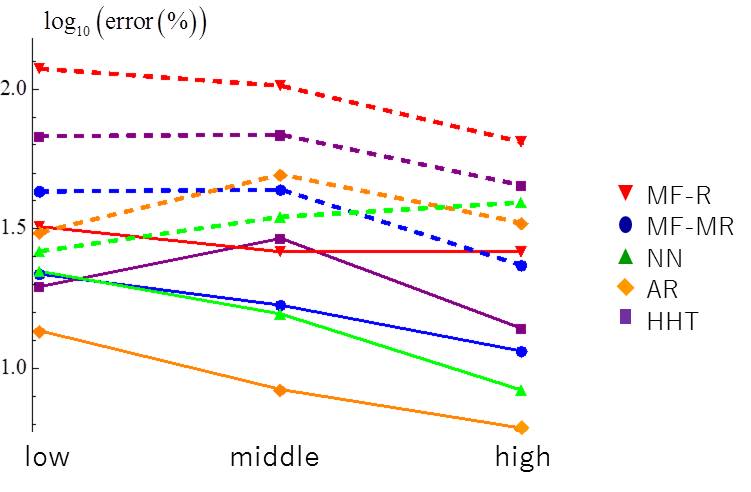}}
   \subfigure[~~Set B only]{
      \includegraphics[width=.48\textwidth]{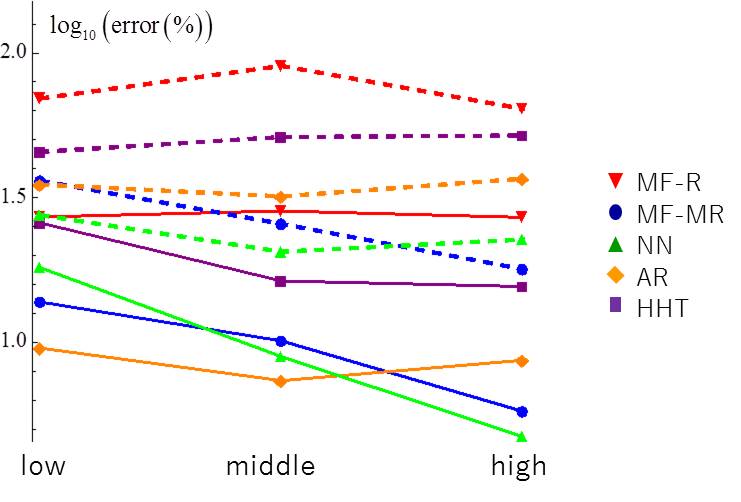}}
\caption{The root mean square error in percent of each method for three levels of SNR. 
The solid and dashed lines represent the real and imaginary parts, respectively.
\label{fig:5}}
  \end{center}
\end{figure*}

To show how the errors depend on SNR, we present several plots of the averaged values 
within each level of SNR, high, middle and low, i.e.,  $\rho_{\rm all}=60,\, 30$ and $20$. 
The variances of the differences $\delta \log f_{\rm R}=\log f_{\rm R}-\log f_{\rm R}^{\rm (inj)}$, and 
$\delta \log f_{\rm I}=\log f_{\rm I}-\log f_{\rm I}^{\rm (inj)}$
are shown in Fig.~\ref{fig:5}, respectively. 
The 
solid lines denote the results for the real part while the dashed lines those for the imaginary part.
The estimations of $f_{\rm I}$ are generally about 0.5-order worse than those of $f_{\rm R}$.  This tells us the difficulty of identifying the damping rate. 
As expected, the differences are smaller for larger $\rho_{\rm all}$, with some exceptions.
The main message we can read from Fig.~\ref{fig:5} combined with Table~\ref{table_listdata} is that we would be able to estimate $f_{\rm R}$ within 7\% (8 \%) from the injected value for the data $\rho_{\rm rd}\sim 15$ $(8)$, if we adopt an appropriate method. 
On the other hand, the estimate of $f_{\rm I}$ has an error at least of $O(30\%)$ even for the data with $\rho_{\rm rd}\sim 15$.

\begin{figure*}[!th]
  \begin{center}
    \includegraphics[width=1\textwidth]{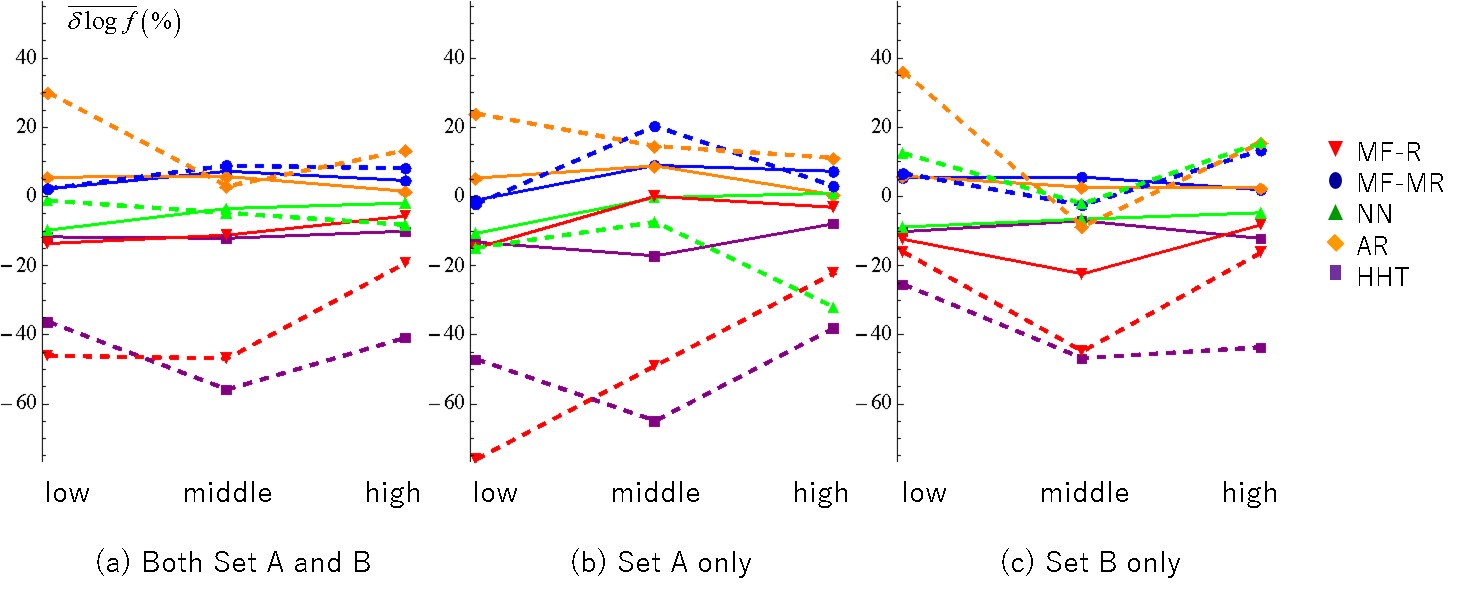}
\caption{The averaged error in percent of each method for three levels of SNR. 
The solid and dashed lines represent the real and imaginary parts, respectively. 
\label{fig:6}}
  \end{center}
\end{figure*}

The averages of $\delta \log f_{\rm R}$ and $\delta\log f_{\rm I}$ are also shown in Fig.~\ref{fig:6}. 
For all five methods, we see the estimated values of $f_{\rm R}$ are roughly distributed around the injected one $f_{\rm R}^{\rm (inj)}$, while there are some tendencies that $f^{\rm (inj)}_{\rm I}$ is over or underestimated, depending on the method.  These results would be suggestive in the interpretation of the future application of each method to real data.

%


\subsubsection{Error estimate}
MF-MR and AR methods give the error estimates 
as explained in Sec.~\ref{section:3}. 
The consistency of these error estimates is briefly checked below. 

In the case of MF-MR method for set B data, the 
expected result is obtained. Namely, there are 120 guesses in the present 
test (60 real parts and 60 imaginary parts). The 90\% confidence interval is  
given by cutting 5\% probability regions on both small and large value sides. 
This estimate of the confidence interval just take into account the 
statistical error. 
The true value fell outside of the confidence interval 27 times out of 120. 
This is slightly worse than the expectation. The estimate of the confidence interval
may need modification. 
For set A data, this happened 45 times, 
which means that the contribution of the systematic bias is significantly large. 

For AR method, the true value becomes outside the 90\% confidence interval 
21 times out of 120 guesses for the imaginary part while it happened 51 times 
for the real part. 

\section{Concluding Remarks}

We implemented five methods for extracting ringdown waves solely, and tested them with mock data by a method of ``blind analysis".  

Comparison tells that AR method, which can pick up the frequency of ringdown wave $f_{\rm R}$ with 10\% root mean square difference from the injected one for the SNR of the ringdown part greater than 7 or so, showed the best performance in determining the real part. AR method is superseded by NN or MF-MR method for Set B data with high SNR of the ringdown part greater than 12 or so. 
The same template as the Set B data is used as the training data for NN method and as the template to be matched for MF-MR method. 
On the other hand, the imaginary part of the frequency $f_{\rm I}$ (related to the damping period) is rather difficult to determine, and  AR, NN and MF-MR methods showed comparable performance. 
The data tells that the root mean square difference of $f_{\rm I}$ from the injected one for high SNR data can be less than about 30\%, although the result would apply only 
in modifications of the ringdown waveform limited to the one smoothly connected to the merger phase. 
We believe that the possibility and the limitation of independent estimation of ringdown mode was shown in this paper, and this opens a way of testing gravity theories.

When the error circles derived by using some combination of several methods are overlapping, we might be able to more confidently claim that the QNM frequency is determined by the observational data.  However, currently only two of our analysis methods (MF-MR, AR) reported error circles, 
and the estimated error circles also contain some errors.
Once we have various methods whose error estimate is reliable, there might be a possibility to combine the estimates properly.

Through the mock data challenges, we also learned the directions of further improvements of each methods.
\begin{itemize}
    \item The MF methods do not have much room for further improvement. As for MF-MR, one possible extension is to adopt a little wider class of templates which depend on parameters other than the QNM frequency. However, the preliminary trial calculations suggest that the extension in this direction will not be so successful.
    
    \item The AR method, presented here, used the Burg method for fitting data and final prediction error method for fixing length of data sequence.  We think that it will be interesting to compare with similar but slightly different approaches, such as those proposed by Berti {\it et al.}~\cite{BCGS2007}. 
    
    \item In the HHT method, there exists the mode-splitting problem of the EMD~\cite{YehAADA2010}. We are planning to resolve the problem by taking into account the sparsity in the frequency domain to the EMD. It may improve the accuracy of the extraction of a QNM since the instantaneous frequency of the QNM is constant.
    
    \item  First, the neural network is trained with waveforms generated by the same method as the set B. So, the results for the set A seems to contain bias in high SNR regime. The improvement of the training algorithm or preparing the dataset will reduce the bias. Second, the NN method can give only the central value for the current estimation.  We need to find a method to estimate the prediction errors.
    
\end{itemize}

After implementations of such improvements, we are planning to apply our methods to the real GW data, to discuss the validity of general relativity.

\begin{acknowledgments}
This research made use of data generated
by the Simulating eXtreme Spacetimes.
This work was supported by JSPS KAKENHI Grant Number JP17H06358 (and also JP17H06357), 
{\it A01: Testing gravity theories using gravitational waves}, as a part of 
the innovative research area, ``Gravitational wave physics and astronomy: Genesis".
H.~N. acknowledges support from JSPS KAKENHI Grant No. JP16K05347. 
T.~N.'s work was also supported in part by a Grant-in-Aid for JSPS Research Fellows.
H.~S. acknowledges also supports from JSPS KAKENHI Grant Nos. JP18K03630 and 19H01901. 
H.~T. acknowledges support from JSPS KAKENHI Grant No. JP17K05437.
T.~T. acknowledges support from JSPS KAKENHI Grant Nos. JP26287044 and JP15H02087.
\end{acknowledgments}


\end{document}